\documentclass[journal]{IEEEtran}
\IEEEoverridecommandlockouts
\usepackage{amsmath}
\usepackage{amsfonts}
\usepackage{amsthm} 
\usepackage{amssymb} 
\usepackage{breqn}
\usepackage{setspace}
\usepackage[usenames,dvipsnames]{xcolor}
\usepackage{graphicx}
\usepackage{endnotes}
\usepackage{mdframed}
       \usepackage{sidecap}
       \sidecaptionvpos{figure}{c}
\usepackage[T1]{fontenc}
\usepackage{supertabular}
\usepackage{longtable}
\usepackage{bbm}
\usepackage{url}
\usepackage{enumerate}
\usepackage{fixltx2e}

\usepackage[normalem]{ulem}

\usepackage{mdwmath}
\usepackage{mdwtab}

\title{\color{Brown} 
On the shadow moments of \\
apparently infinite-mean phenomena
}

\author{
    \IEEEauthorblockN{Nassim Nicholas Taleb\IEEEauthorrefmark{1} and Pasquale Cirillo\IEEEauthorrefmark{2}}\\
   
     \IEEEauthorblockA{\IEEEauthorrefmark{1}School of Engineering, New York University\\}
    \IEEEauthorblockA{\IEEEauthorrefmark{2}Applied Probability Group, Delft University of Technology}
}

\date{}

\begin{document}

\maketitle

\setstretch{1.1}
\begin{abstract}\small{
We propose an approach to compute the conditional moments of fat-tailed phenomena that, only looking at data, could be mistakenly considered as having infinite mean. This type of problems manifests itself when a random variable $Y$ has a heavy-tailed distribution with an extremely wide yet bounded support.

We introduce the concept of dual distribution, by means of a log-transformation that smoothly removes the upper bound. The tail of the dual distribution can then be studied using extreme value theory, without making excessive parametric assumptions, and the estimates one obtains can be used to study the original distribution and compute its moments by reverting the transformation.

The central difference between our approach and a simple truncation is in the smoothness of the transformation between the original and the dual distribution, allowing use of extreme value theory.

War casualties, operational risk, environment blight, complex networks and many other econophysics phenomena are possible fields of application.}

\end{abstract} 
\maketitle 

\section{Introduction}
Consider a heavy-tailed random variable $Y$ with finite support $[L,H]$. W.l.o.g. set $L>>0$ for the lower bound, while for upper one $H$, assume that its value is remarkably large, yet finite. It is so large that the probability of observing values in its vicinity is extremely small, so that in data we tend to find observations only below a certain $M<<H<\infty$.

\begin{figure}[!htb]
\includegraphics[width=\linewidth]{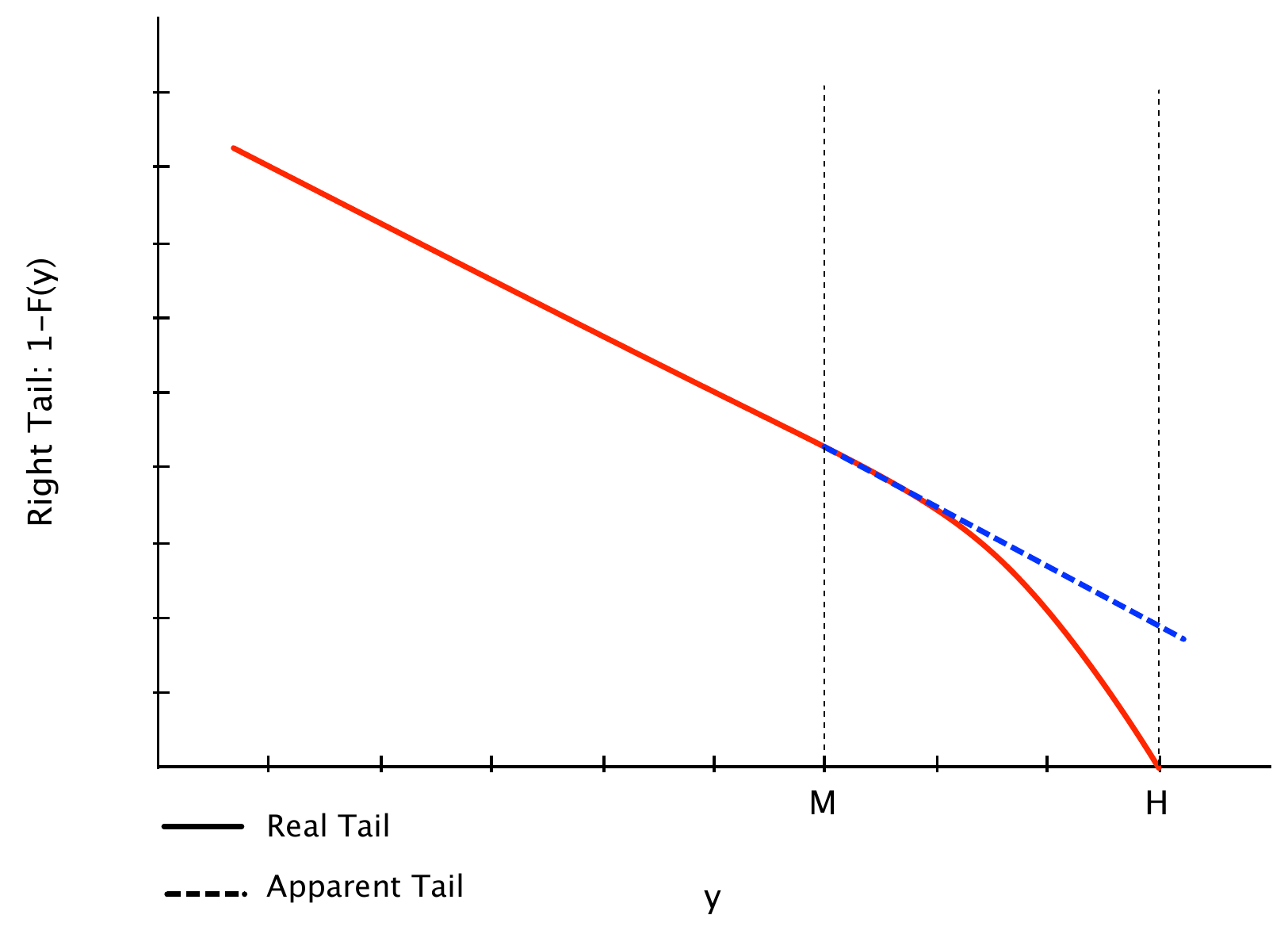}
\caption{Graphical representation of what may happen if one ignores the existence of the finite upper bound $H$, since only $M$ is observed.}
\label{tailcomp}
\end{figure}

Figure \ref{tailcomp} gives a graphical representation of the problem. For our random variable $Y$ with remote upper bound $H$ the real tail is represented by the continuous line. However, if we only observe values up to $M<<H$, and - willing or not - we ignore the existence of $H$, which is unlikely to be seen, we could be inclined to believe the the tail is the dotted one, the apparent one. The two tails are indeed essentially indistinguishable for most cases, as the divergence is only evident when we approach $H$. 

Now assume we want to study the tail of $Y$ and, since it is fat-tailed and despite $H<\infty$, we take it to belong to the so-called Fr\'echet class\footnote{Note that treating $Y$ as belonging to the Fr\'echet class is a mistake. If a random variable has a finite upper bound, it cannot belong to the Fr\'echet class, but rather to the Weibull class \cite{deHaan}.}. In extreme value theory \cite{Embrechts}, a distribution $F$ of a random variable $Y$ is said to be in the Fr\'echet class if $\bar{F}(y)=1-F(y)=y^{-\alpha}L(y)$, where $L(y)$ is a slowly varying function. In other terms, the Fr\'echet class is the class of all distributions whose right tail behaves as a power law.

Looking at the data, we could be led to believe that the right tail is the dotted line in Figure \ref{tailcomp}, and our estimation of $\alpha$ shows it be smaller than 1. Given the properties of power laws, this means that $E[Y]$ is not finite (as all the other higher moments). This also implies that the sample mean is essentially useless for making inference, in addition to any considerations about robustness \cite{Maronna}. But if $H$ is finite, this cannot be true: all the moments of a random variable with bounded support are finite.

A solution to this situation could be to fit a parametric model, which allows for fat tails and bounded support, such as for example a truncated Pareto \cite{Inmaculada}. But what happens if $Y$ only shows a Paretian behavior in the upper tail, and not for the whole distribution? Should we fit a mixture model?

In the next section we propose a simple general solution, which does not rely on strong parametric assumptions.

\section{The dual distribution}
Instead of altering the tails of the distribution we find it more convenient to transform the data and rely on distributions with well known properties. In Figure \ref{tailcomp}, the real and the apparent tails are indistinguishable to a great extent. We can use this fact to our advantage, by transforming $Y$ to remove its upper bound $H$, so that the new random variable $Z$ - the dual random variable - has the same tail as the apparent tail. We can then estimate the shape parameter $\alpha$ of the tail of $Z$ and come back to $Y$ to compute its moments or, to be more exact, to compute its excess moments, the conditional moments above a given threshold, view that we will just extract the information from the tail of $Z$.

Take $Y$ with support $[L,H]$, and define the function
\begin{equation}
\varphi(Y)= 
L-H \log \left(\frac{H-Y}{H-L}\right).
\label{transform}
\end{equation}
We can verify that $\varphi$  is "smooth": $\varphi \in C^\infty$, $\varphi^{-1}(\infty)=H$, and $\varphi^{-1}(L)=\varphi(L)=L$. Then $Z=\varphi(Y)$ defines a new random variable with lower bound $L$ and an infinite upper bound. Notice that the transformation induced by $\varphi(\cdot)$ does not depend on any of the parameters of the distribution of $Y$. 

By construction, $z=\varphi(y)\approx y$ for very large values of $H$. This means that for a very large upper bound, unlikely to be touched, the results we get for the tail of $Y$ and $Z=\varphi(Y)$ are essentially the same, until we do not reach $H$. But while $Y$ is bounded, $Z$ is not. Therefore we can safely model the unbounded dual distribution of $Z$ as belonging to the Fr\'echet class, study its tail, and then come back to $Y$ and its moments, which under the dual distribution of $Z$ could not exist.\footnote{Note that the use of logarithmic transformation is quite natural in the context of utility.}

The tail of $Z$ can be studied in different ways, see for instance \cite{Embrechts} and \cite{Falk}. Our suggestions is to rely on the so-called de Pickands, Balkema and de Haan's Theorem \cite{deHaan}. This theorem allows us to focus on the right tail of a distribution, without caring too much about what happens below a given threshold threshold $u$. In our case $u\geq L$.

Consider a random variable $Z$ with distribution function $G$, and call $G_u$ the conditional df of $Z$ above a given threshold $u$. We can then define the r.v. $W$, representing the rescaled excesses of $Z$ over the threshold $u$, so that
$$G_u(w) = P(Z-u \leq w | Z>u) = \frac{G(u+w)-G(u)}{1-G(u)}, $$
for $0 \leq w \leq z_G-u$, where $z_G$ is the right endpoint of $G$. 

Pickands, Balkema and de Haan have showed that for a large class of distribution functions $G$, and a large $u$, $G_u$ can be approximated by a Generalized Pareto distribution, i.e.
$G_u(w) \rightarrow GPD(w; \xi, \sigma),\text{ as }u \rightarrow \infty$
where
\begin{equation}
GPD(w; \xi, \sigma)=\begin{cases}
1-(1+\xi \frac{w}{\sigma})^{-1/\xi} &  if \,\xi \neq 0\\
 1-e^{-\frac{w}{\sigma}} &  if \,\xi = 0
\end{cases}, \; w\geq 0.
\label{GPD}
\end{equation}
The parameter $\xi$, known as the shape parameter, and corresponding to $1/\alpha$, governs the fatness of the tails, and thus the existence of moments. The moment of order $p$ of a Generalized Pareto distributed random variable only exists if and only if $\xi < 1/p$, or $\alpha > p$ \cite{Embrechts}. Both $\xi$ and $\sigma$ can be estimated using MLE or the method of moments \cite{deHaan}.\footnote{There are alternative methods to face finite (or concave) upper bounds, i.e., the use of tempered power laws (with exponential dampening)\cite{Rachev} or stretched exponentials \cite{Laherrere}; while being of the same nature as our exercise, these methods do not allow for immediate applications of extreme value theory or similar methods for parametrization.}

\section{Back to $Y$: the shadow mean}
With $f$ and $g$, we indicate the densities of $Y$ and $Z$.

We know that $Z=\varphi(Y)$, so that $Y=\varphi^{-1}(Z)=(L-H) e^{\frac{L-Z}{H}}+H$. 

Now, let's assume we found $u=L^\ast \geq L$, such that $G_u(w)\approx \text{GPD}(w;\xi,\sigma)$. This implies that the tail of $Y$, above the same value $L^\ast$ that we find for $Z$, can be obtained from the tail of $Z$, i.e. $G_u$.

First we have
\begin{equation}
\int_{L^\ast}^\infty g(z) \, \mathrm{d} z = \int_{L^\ast}^{\varphi^{-1}(\infty)} f(y) \, \mathrm{d} y.
\end{equation}
And we know that
\begin{equation}
g(z;\xi,\sigma)=\frac{1}{\sigma} \left(1+\frac{\xi z}{\sigma}\right)^{-\frac{1}{\xi}-1},\qquad z\in [{L^\ast},\infty).
\end{equation}
Setting $\alpha=\xi^{-1}$, we get
\begin{equation}\label{df}
f(y;\alpha,\sigma)=\frac{H \left(1+\frac{H (\log (H-L)-\log (H-y))}{\alpha  \sigma }\right)^{-\alpha -1}}{\sigma  (H-y)}, \;\; y \in [{L^\ast},H],
\end{equation}
or, in terms of distribution function,
\begin{equation}\label{DF}
F(y;\alpha,\sigma)=1-\left(1+\frac{H (\log (H-L)-\log (H-y))}{\alpha  \sigma }\right)^{-\alpha }.
\end{equation}
Clearly, given that $\varphi$ is a one-to-one transformation, the parameters of $f$ and $g$  obtained by maximum likelihood methods will be the same ---the likelihood functions of $f$ and $g$ differ by a scaling constant.

We can  derive the shadow mean\footnote{We call it "shadow", as it is not immediately visible from the data.} of $Y$, conditionally on $Y>L^\ast$, as
\begin{equation}
E[Y|Y>L^\ast]=\int_{L^\ast}^H y \, f(y;\alpha,\sigma) \,\mathrm{d} y ,
\end{equation}
obtaining
\begin{equation}\label{meanY}
E[Y|Z>L^\ast]=(H-L^\ast) e^{\frac{\alpha  \sigma }{H}} \left(\frac{\alpha  \sigma }{H}\right)^{\alpha } \Gamma \left(1-\alpha ,\frac{\alpha  \sigma }{H}\right)+L^\ast. 	
\end{equation}

The conditional mean of $Y$ above $L^\ast\geq L$ can then be estimated by simply plugging in the estimates $\hat{\alpha}$ and $\hat{\sigma}$, as resulting from the GPD approximation of the tail of $Z$. It is worth noticing that if $L^\ast=L$, then $E[Y|Y>L^\ast]=E[Y]$, i.e. the conditional mean of $Y$ above $Y$ is exactly the mean of $Y$.

Naturally, in a similar way, we can obtain the other moments, even if we may need numerical methods to compute them.

Our method can be used in general, but it is particularly useful when, from data, the tail of $Y$ appears so fat that no single moment is finite, as it is often the case when dealing with operational risk losses, the degree distribution of large complex networks, or other econophysical phenomena.

For example, assume that for $Z$ we have $\xi>1$. Then both $E[Z|Z>L^\ast]$ and $E[Z]$ are not finite\footnote{Remember that for a GPD random variable $Z$, $E\left[Z^ p\right]<\infty$ iff $\xi < 1/p$.}. Figure \ref{tailcomp} tells us that we might be inclined to assume that also $E[Y]$ is infinite - and this is what the data are likely to tell us if we estimate $\hat{\xi}$ from the tail\footnote{Because of the similarities between $1-F(y)$ and $1-G(z)$, at least up until $M$, the GPD approximation will give two statistically undistinguishable estimates of $\xi$ for both tails \cite{Embrechts}.} of $Y$. But this cannot be true because $H<\infty$, and even for $\xi>1$ we can compute the expected value $E[Y|Z>L^\ast]$ using equation (\ref{meanY}).

\subsection*{Value-at-Risk and Expected Shortfall}
Thanks to equation (\ref{DF}), we can compute by inversion the quantile function of $Y$ when $Y\geq L^\ast$, that is
\begin{equation}\label{QF}
Q(p;\alpha,\sigma, H, L)=e^{-\gamma(p)} \left(L^\ast e^{\frac{\alpha  \sigma }{H}}+H e^{\gamma(p)} -H e^{\frac{\alpha  \sigma }{H}}\right),
\end{equation}
where $\gamma(p)=\frac{\alpha  \sigma  (1-p)^{-1/\alpha }}{H}$ and $p\in[0,1]$. Again, this quantile function is conditional on $Y$ being larger than $L^\ast$. 

From equation (\ref{QF}), we can easily compute the  Value-at-Risk ($VaR$) of $Y|Y\geq L^\ast$ for whatever confidence level. For example, the $95\%$ $VaR$ of $Y$, if $Y$ represents operational losses over a 1-year time horizon, is simply $VaR_{0.95}^Y=Q(0.95;\alpha,\sigma, H, L)$.

Another quantity we might be interested in when dealing with the tail risk of $Y$ is the so-called expected shortfall ($ES$), that is $E[Y|Y>u\geq L^\ast]$. This is nothing more than a generalization of equation (\ref{meanY}).

We can obtain the expected shortfall by first computing the mean excess function of $Y|Y\geq L^\ast$, defined as
$$
e_u(Y)=E[Y-u|Y>u]=\frac{\int_{u}^\infty (u - y) f(y;\alpha,\sigma)\text{d}y}{1-F(u)},
$$
for $y\geq u \geq L^\ast$. Using equation (\ref{df}), we get
\begin{eqnarray} \label{me}
e_u(Y)&=&(H-L) e^{\frac{\alpha  \sigma }{H}} \left(\frac{\alpha  \sigma }{H}\right)^{\alpha } \left(\frac{H \log \left(\frac{H-L}{H-u}\right)}{\alpha  \sigma }+1\right)^{\alpha } \times \nonumber \\
&&\Gamma \left(1-\alpha ,\frac{\alpha  \sigma }{H}+\log \left(\frac{H-L}{H-u} \right)\right).
\end{eqnarray}
The Expected Shortfall is then simply computed as
$$
E[Y|Y>u\geq L^\ast]=e_u(Y)+u.
$$
As in finance and risk management, ES and VaR can be combined. For example we could be interested in computing the $95\%$ ES of $Y$ when $Y\geq L^\ast$. This is simply given by $VaR_{0.95}^Y+e_{VaR_{0.95}^Y}(Y)$.
\section{Comparison to other methods}
There are three ways to go about explicitly cutting a Paretan distribution in the tails (not counting methods to stretch or "temper" the distribution). 

1) The first one consists in hard truncation, i.e. in setting a single endpoint for the distribution and normalizing. For instance the distribution would be normalized between $L$ and $H$, distributing the excess mass across all points. 

2) The second one would assume that $H$ is an absorbing barrier, that all the realizations of the random variable in excess of $H$ would be compressed into a Dirac delta function at $H$ --as practiced in derivative models. In that case the distribution would have the same density as a regular Pareto except at point $H$.

3) The third is the one presented here.

The same problem has cropped up in quantitative finance over the use of truncated normal (to correct for Bachelier's use of a straight Gaussian) vs. logarithmic transformation (Sprenkle, 1961 \cite{sprenkle}), with the standard model opting for logarithmic transformation and the associated one-tailed lognormal distribution. Aside from the additivity of log-returns and other such benefits, the models do not produce a "cliff", that is an abrupt change in density below or above, with the instability associated with risk measurements on non-smooth function.   

\begin{figure}[!htb]
\includegraphics[width=\linewidth]{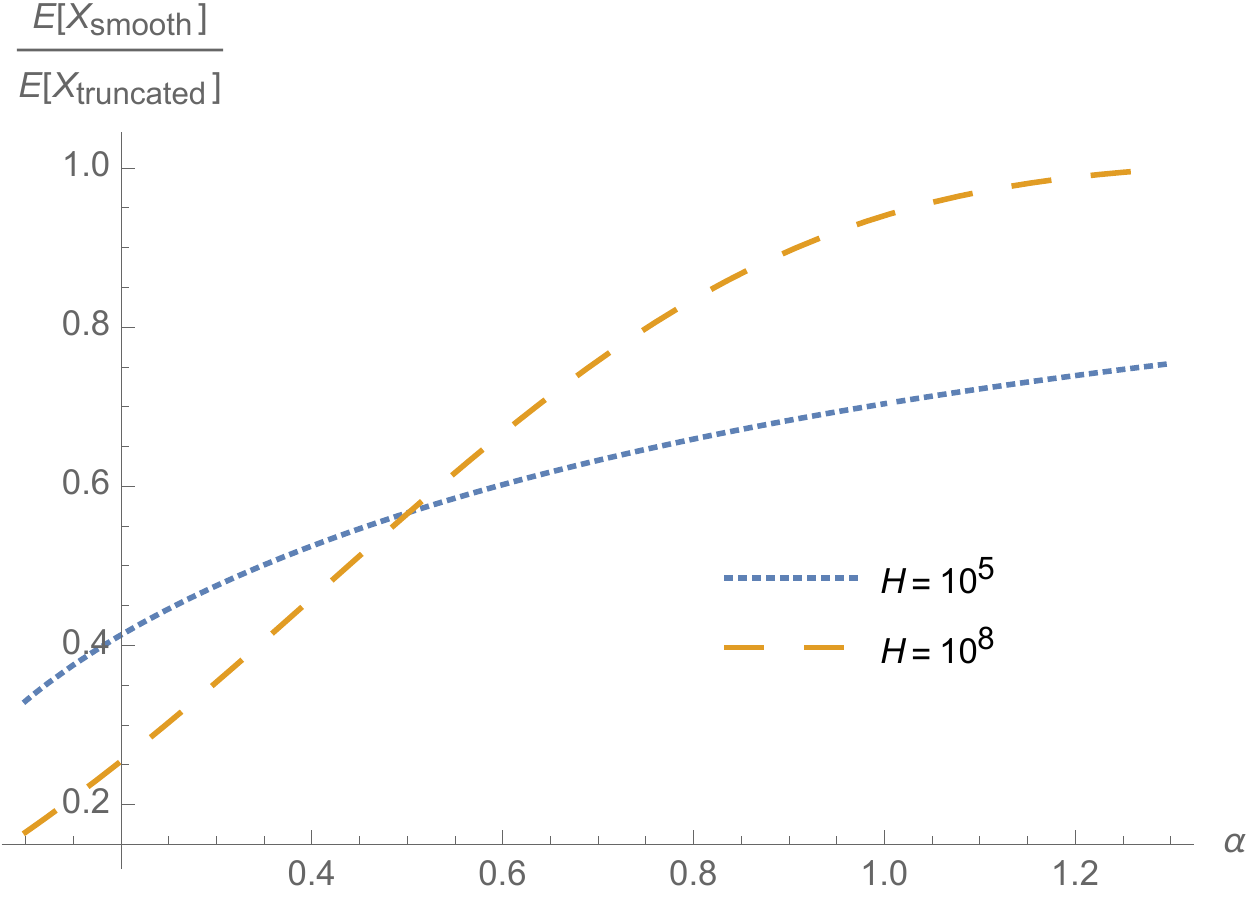}
\label{ratio}
\caption{Ratio of the expectation of smooth transformation to truncated.}
\end{figure}

As to the use of extreme value theory, Breilant et al. (2014)\cite{breilant} go on to truncate the distribution by having an excess in the tails with the transformation $Y^{-\alpha} \rightarrow (Y^{-\alpha} -H^{-\alpha})$ and apply EVT to the result. Given that the transformation includes the estimated parameter, a new MLE for the parameter $\alpha$ is required. We find issues with such a non-smooth transformation. The same problem occurs as with financial asset models, particularly the presence an abrupt "cliff" below which there is a density, and above which there is none. The effect is that the expectation obtained in such a way will be higher than ours, particularly at values of $\alpha < 1$, as seen in Figure \ref{ratio}.

We can demonstrate the last point as follows. Assume we observe distribution is Pareto that is in fact truncated but treat it as a Pareto.
The density is $f(x)=\frac{1}{\sigma}\left(\frac{x-L}{\alpha  \sigma }+1\right)^{-\alpha -1}\, , x \in [L,\infty)$. The truncation gives $ g(x)=\frac{\left(\frac{x-L}{\alpha  \sigma }+1\right)^{-\alpha -1}}{\sigma  \left(1-\alpha ^{\alpha } \sigma ^{\alpha } (\alpha  \sigma +H-L)^{-\alpha }\right)} \, , x \in [L,H]$.

Moments of order $p$ of the truncated Pareto (i.e. what is seen from realizations of the process), $M(p)$ are:
\begin{equation}
\begin{aligned}
	M(p)=&\alpha  e^{-i \pi  p} (\alpha  \sigma )^{\alpha } (\alpha  \sigma -L)^{p-\alpha }\\
	&\frac{\left(B_{\frac{H}{L-\alpha  \sigma }}(p+1,-\alpha )-B_{\frac{L}{L-\alpha  \sigma }}(p+1,-\alpha )\right)}{\left(\frac{\alpha  \sigma }{\alpha  \sigma +H-L}\right)^{\alpha }-1}
\end{aligned}
\end{equation}
where $B(.,.)$ is the Euler Beta function, $B(a,b)=\frac{ \Gamma(a )  \Gamma (b)}{\Gamma  (a+b)}=\int _0^1 t^{a-1} (1-t)^{b-1} \,dt$.

We end up with $r(H,\alpha)$, the ratio of the mean of the soft truncated to that of the truncated Pareto.
\begin{equation}
\begin{aligned}
	r(H,\alpha)=&e^{-\frac{\alpha }{H}} \left(\frac{\alpha }{H}\right)^{\alpha } \left(\frac{\alpha }{\alpha +H}\right)^{-\alpha } \left(\frac{\alpha +H}{\alpha }\right)^{-\alpha } \\
	&\frac{\left(-\left(\frac{\alpha +H}{\alpha }\right)^{\alpha }+H+1\right)}{(\alpha -1) \left(\left(\frac{\alpha }{H}\right)^{\alpha }-\left(\frac{\alpha +H}{H}\right)^{\alpha }\right) E_{\alpha }\left(\frac{\alpha }{H}\right)}	
	\end{aligned}
\end{equation}
where $E_{\alpha }\left(\frac{\alpha }{H}\right)$ is the exponential integral $e_\alpha z=\int _1^{\infty }\frac{e^{t (-\alpha)}}{t^n}\, dt$.

\section{Applications}
\subsubsection*{Operational risk} The losses for a firm are bounded by the capitalization, with well-known maximum losses.
\subsubsection*{Capped Reinsurance Contracts} Reinsurance contracts almost always have caps (i.e., a maximum claim); but a reinsurer can have many such contracts on the same source of risk and the addition of the contract pushes the upper bound in such a way as to cause larger potential cumulative harm.
\subsubsection*{Violence} While wars are extremely fat-tailed, the maximum effect from any such event cannot exceed the world's population.
\subsubsection*{Credit risk} A loan has a finite maximum loss, in a way similar to reinsurance contracts.
\subsubsection*{City size} While cities have been shown to be Zipf distributed, the size of a given city cannot exceed that of the world's population.
\subsubsection*{Environmental harm} While these variables are exceedingly fat-tailed, the risk is confined by the size of the planet (or the continent on which they take place) as a firm upper bound.

\subsubsection*{Complex networks} The number of connections is finite.
\subsubsection*{Company size} The sales of a company is bound by the GDP.
\subsubsection*{Earthquakes} The maximum harm from an earthquake is bound by the energy.
\subsubsection*{Hydrology} The maximum level of a flood can be determined.

%
%
%
%


\begin{thebibliography}{99}
\bibitem{breilant} Beirlant, J., Alves, I. F., Gomes, I., and Meerschaert, M. (2014). Extreme value statistics for truncated Pareto-type distributions. arXiv preprint arXiv:1410.4097.   
\bibitem{deHaan} L. de Haan, A. Ferreira (2006). \textit{Extreme Value Theory: An Introduction}. Springer.
\bibitem{Embrechts} P. Embrechts, C. Kl\"uppelberg, T. Mikosch (2003). \textit{Modelling Extremal Events}. Springer.
\bibitem{Falk} M. Falk, J. H\"usler J, R. D. Reiss R-D (2004). \textit{Laws of small numbers: extremes and rare events}, Birkh\"auser.
\bibitem{Inmaculada} B.A. Inmaculada, M.M. Meerschaert, A.K. Panorska (2006). Parameter estimation for the truncated Pareto distribution. \textit{Journal of the American Statistical Association} 101, 270-277.
\bibitem{Maronna} R. Maronna, R.D. Martin, V. Yohai (2006). \textit{Robust Statistics - Theory and Methods}. Wiley.
\bibitem{Rachev} Rachev, S. T., Kim, Y. S., Bianchi, M. L., and Fabozzi, F. J. (2011). \textit{Financial models with L\'evy processes and volatility clustering}. Wiley.
\bibitem{Laherrere} Laherrere, J., and Sornette, D. (1998). Stretched exponential distributions in nature and economy:"fat tails" with characteristic scales. The European Physical Journal B-Condensed Matter and Complex Systems, 2(4), 525-539.
\bibitem{sprenkle} Sprenkle, C., 1961. Warrant prices as indicators of expectations and preferences. \textit{Yale Economics Essays} 1, 179-231.

\bibitem{Taleb} N. N. Taleb (2015). \textit{Silent Risk: Lectures on Probability}, Vol 1. SSRN.
\end{thebibliography}
\end{document}